\documentclass[aps,pra,superscriptaddress,twocolumn,10pt,a4paper]{revtex4-1}
\usepackage{graphicx}
\pdfoutput=1

\begin{document}

\title{Storing quantum information for 30 seconds in a nanoelectronic device}

\author{Juha T. Muhonen}
\affiliation{Centre for Quantum Computation and Communication Technology, School of Electrical Engineering and Telecommunications, University of New South Wales, Sydney, New South Wales 2052, Australia}

\author{Juan P. Dehollain}
\affiliation{Centre for Quantum Computation and Communication Technology, School of Electrical Engineering and Telecommunications, University of New South Wales, Sydney, New South Wales 2052, Australia}

\author{Arne Laucht}
\affiliation{Centre for Quantum Computation and Communication Technology, School of Electrical Engineering and Telecommunications, University of New South Wales, Sydney, New South Wales 2052, Australia}

\author{Fay E. Hudson}
\affiliation{Centre for Quantum Computation and Communication Technology, School of Electrical Engineering and Telecommunications, University of New South Wales, Sydney, New South Wales 2052, Australia}

\author{Takeharu Sekiguchi}
\affiliation{School of Fundamental Science and Technology, Keio University, 3-14-1 Hiyoshi, 223-8522, Japan}

\author{Kohei M. Itoh}
\affiliation{School of Fundamental Science and Technology, Keio University, 3-14-1 Hiyoshi, 223-8522, Japan}

\author{David N. Jamieson}
\affiliation{Centre for Quantum Computation and Communication Technology, School of Physics, University of Melbourne, Melbourne, Victoria 3010, Australia}

\author{Jeffrey C. McCallum}
\affiliation{Centre for Quantum Computation and Communication Technology, School of Physics, University of Melbourne, Melbourne, Victoria 3010, Australia}

\author{Andrew S. Dzurak}
\affiliation{Centre for Quantum Computation and Communication Technology, School of Electrical Engineering and Telecommunications, University of New South Wales, Sydney, New South Wales 2052, Australia}

\author{Andrea Morello}
\affiliation{Centre for Quantum Computation and Communication Technology, School of Electrical Engineering and Telecommunications, University of New South Wales, Sydney, New South Wales 2052, Australia}

\maketitle

\textbf{
The spin of an electron or a nucleus in a semiconductor \cite{Awschalom2013} naturally implements the unit of quantum information -- the qubit -- while providing a technological link to the established electronics industry \cite{Zwanenburg2013}. The solid-state environment, however, may provide deleterious interactions between the qubit and the nuclear spins of surrounding atoms \cite{Yao2006}, or charge and spin fluctuators in defects, oxides and interfaces \cite{Desousa2007}. For group IV materials such as silicon, enrichment of the spin-zero $^{28}$Si isotope drastically reduces spin-bath decoherence \cite{Witzel2010}. Experiments on bulk spin ensembles in $^{28}$Si crystals have indeed demonstrated extraordinary coherence times \cite{Tyryshkin2012,Steger2012,Saeedi2013}. However, it remained unclear whether these would persist at the single-spin level, in gated nanostructures near amorphous interfaces. Here we present the coherent operation of individual $^{31}$P electron and nuclear spin qubits in a top-gated nanostructure, fabricated on an isotopically engineered $^{28}$Si substrate. We report new benchmarks for coherence time ($> 30$~seconds) and control fidelity ($>99.99$\%) of any single qubit in solid state, and perform a detailed noise spectroscopy \cite{Alvarez2011} to demonstrate that -- contrary to widespread belief -- the coherence is not limited by the proximity to an interface. Our results represent a fundamental advance in control and understanding of spin qubits in nanostructures.
}

It is well known that the Si/SiO$_2$ interface hosts a variety of defects that act as charge and spin fluctuators. Spin resonance experiments have documented the deleterious effects of the Si/SiO$_2$ interface on the coherence of donors in $^{28}$Si, implanted at different depths \cite{Schenkel2006}. Theoretical models suggest that magnetic fluctuation from paramagnetic spins at the interface cause the decohering noise \cite{Desousa2007}, and recent work advocates the use of `clock transitions' in $^{209}$Bi donors \cite{Wolfowicz2013} to obtain a spin qubit that is to first-order insensitive to magnetic noise. Fluctuations of interface charges or gate voltages can also cause decoherence, if there is a physical mechanism for electric fields to couple to the spin qubit states. Evidence of such effects was found for instance in carbon nanotube valley-spin qubits \cite{Laird2013}. For donors in silicon, fluctuating electric fields can couple to the spin states by modulating the hyperfine coupling \cite{Rahman2007,Mohiyaddin2013} or the $g$-factor \cite{Rahman2009}. Here we operate single-atom spin qubits in isotopically purified $^{28}$Si, with a residual $^{29}$Si concentration of 800 ppm. Minimizing the effect $^{29}$Si nuclear spin fluctuations allowed us not only to set new benchmarks for qubit performance in solid state, but also to uncover the microscopic origin of residual decoherence mechanisms, specific to a gated nanostructure.

A substitutional P atom in Si behaves to a good approximation like hydrogen in vacuum, with energy levels renormalized by the effective mass and the dielectric constant of the host material \cite{Greenland2010}. Both the bound electron (e$^-$) and the nucleus ($^{31}$P) possess a spin $1/2$ and constitute natural qubits with simple spin up/down eigenstates, which we denote as $|{\uparrow\rangle},|{\downarrow\rangle}$ for e$^-$ and $|{\Uparrow\rangle},|{\Downarrow\rangle}$ for $^{31}$P. The contact hyperfine interaction $A$ between e$^-$ and $^{31}$P, and the application of a static magnetic field $B_0 > 1$~T result in a 4-level energy diagram as shown in Fig.~1\textbf{c}. At high magnetic fields the eigenstates are, to a very good approximation, the separable tensor products of the electro-nuclear basis states.

The device structure is shown in Fig.~1\textbf{a,b}. It consists of a silicon single-electron transistor (SET) for spin readout \cite{Morello2010}, a broadband on-chip microwave antenna to deliver an oscillating magnetic field $B_1$ to the qubits \cite{Dehollain2013}, ion-implanted P donors \cite{Jamieson2005}, and a stack of aluminum gates above the SiO$_2$ insulator to control the potentials of the donors and the SET. All the data presented here are obtained from the analysis of single-shot electron \cite{Morello2010} and nuclear \cite{Pla2013} spin readout events. We have measured two devices, A and B, which differ slightly in their gate layout and ion-implantation parameters (see methods).  The experiments were performed in high magnetic fields ($B_0 = 1.62$~T for device A, $B_0 = 1.5$~T for device B) and low temperatures (electron temperature $T_{\rm{el}} \approx 100$~mK).
The two devices had significantly different hyperfine constants ($A/h \approx 116.6$ MHz for device A and 96.9 MHz for device B), probably resulting from a combination of different donor depths, electric fields \cite{Rahman2007,Mohiyaddin2013} or strain \cite{Huebl2006}. We made no attempt to actively tune $A$, but we note that the observed difference corresponds to $>10,000$~times the linewidth of the spin resonance transitions (see below). Engineering and controlling $A$ over the observed range would therefore allow very precise individual addressing of individual qubits in a large register.

\begin{figure}
\centering
\includegraphics[width=0.47\textwidth]{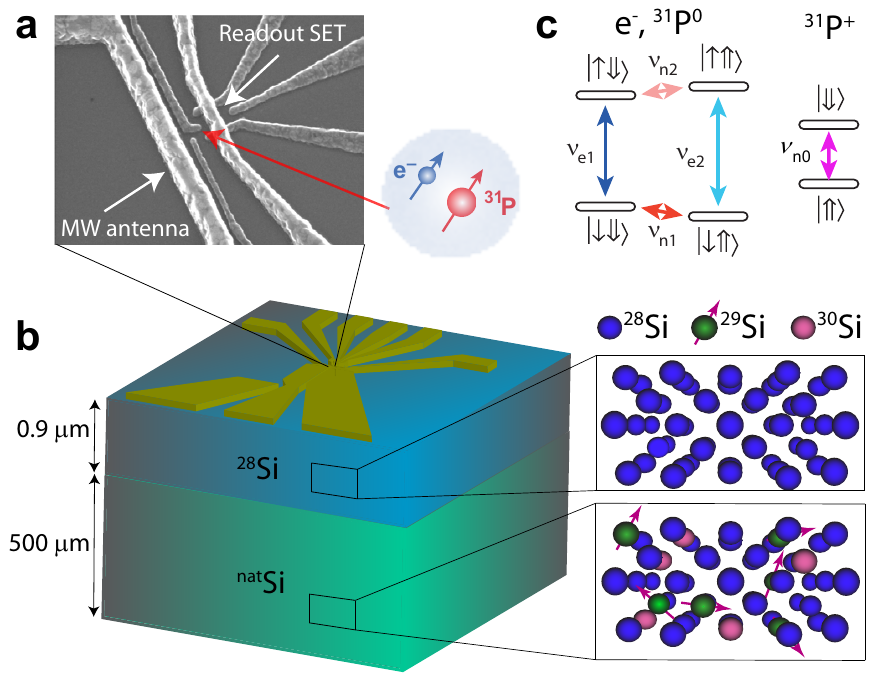}
\caption{\textbf{Device structure and electron/nuclear spin qubits}. \textbf{a}, Scanning electron micrograph image of a device similar to Device A, highlighting the position of the P donor, the microwave (MW) antenna, and the SET for spin readout. \textbf{b}, Schematic of the Si substrate, consisting of an isotopically purified $^{28}$Si epilayer (with a residual $^{29}$Si concentration of 800 ppm) on top of a natural Si wafer. \textbf{c}, Energy level diagram of the coupled e$^-$--$^{31}$P$^0$ system (left) and the ionized $^{31}$P$^+$ nucleus (right). Arbitrary quantum states are encoded on the qubits by applying pulses of oscillating magnetic field $B_1$ at the frequencies corresponding to the electron spin resonance (ESR), $\nu_{e1,2} \approx \gamma_e B_0 \pm A/2$, and nuclear magnetic resonance (NMR), $\nu_{n1,2} \approx A/2 \pm \gamma_n B_0$, where $\gamma_e = 27.97$~GHz/T and $\gamma_n = 17.23$~MHz/T are the electron and nuclear gyromagnetic ratios, respectively. The $^{31}$P qubit in the ionized state is operated at the frequency $\nu_{n0} = \gamma_n B_0$.
}
\label{fig:fig1}
\end{figure}

We report a complete set of qubit control and coherence benchmarks, that include: (i) Rabi oscillations, to prepare arbitrary superposition states of the qubit; (ii) Ramsey fringes, which yield the pure dephasing time $T_2^*$; (iii) Hahn echoes, which yield the qubit coherence time $T_2^{\rm{H}}$; (iv) Carr-Purcell-Meiboom-Gill (CPMG$_N$) dynamical decoupling sequences, used here both to measure the ultimate limit of the coherence time $T_2^{\rm DD}$ and to extract the spectrum of the noise that couples to the qubits (see supplementary section B for details).

The coherent operation of the e$^-$ qubit is shown in Fig.~2\textbf{a}. The Rabi oscillations continue for over 500 $\mu$s before any signs of decay. This is a tremendous improvement over the e$^-$ qubit in $^{\rm{nat}}$Si, where the Rabi oscillations decayed in less than 1~$\mu$s \cite{Pla2012}. The Ramsey experiment (Fig.~2\textbf{b}) yields a pure electron dephasing time $T_{2e}^* = 270$~$\mu$s on Device A -- a 5,000-fold improvement over the $^{\rm{nat}}$Si value of 55~ns, and comparable to the values obtained with nitrogen-vacancy (NV) electron spins in isotopically purified $^{12}$C diamond \cite{Bala2009,Maurer2012}. The corresponding full-width half maximum of the electron spin resonance (ESR) linewidth is $\Delta\nu_{\textsc{fwhm}} = 1 / (\pi T_{2e}^*) = 1.2$~kHz (see supplementary section C for direct measurement of linewidths). With a Hahn echo sequence we measured electron coherence times $T_{2e}^{\rm H} \approx 1$~ms in both devices (Fig.~2\textbf{c}), only a factor 5 longer than in $^{\rm{nat}}$Si \cite{Pla2012}. However, using the CPMG dynamical decoupling technique we extended the e$^-$ spin coherence of the order of 1 second, $T_{2e}^{\rm DD} = 0.56$~s in Device B (Fig.~4\textbf{a}).

For the $^{31}$P qubit we report coherence measurements in the neutral ($^{31}$P$^0$) and the ionized ($^{31}$P$^+$) case (Fig.~2\textbf{c,d}). The $^{31}$P$^0$ shows a similar dephasing time to e$^-$, $T_{2n0}^* \approx 500$~$\mu$s. The Hahn echo decay was found to be very different between Devices A and B, with values 1.5~ms and 20~ms, respectively. As observed before in both single-atom \cite{Pla2013} and bulk experiments \cite{Saeedi2013}, the nuclear spin coherence improves dramatically by removing the electron from the P atom. The $^{31}$P$^+$ Ramsey decay times reached the value $T_{2n+}^* = 0.6$~s in Device B, which would correspond to an NMR linewidth $\Delta\nu_{\textsc{fwhm}} \approx 0.5$~Hz. The simple Hahn echo sequence preserves the qubit coherence beyond 1 second, $T_{2n+}^{\rm H} = 1.75$~s, and the CPMG dynamical decoupling extends it beyond 30 s, $T_{2n+}^{\rm DD} = 35.6$~s in Device B (Fig.~4\textbf{a}). This currently represents the record coherence for any \emph{single} qubit in solid state. A summary of the coherence benchmarks for e$^-$, $^{31}$P$^0$ and $^{31}$P$^+$ in both devices is shown in the supplementary section A.

\begin{figure}
\centering
\includegraphics[width=0.47\textwidth]{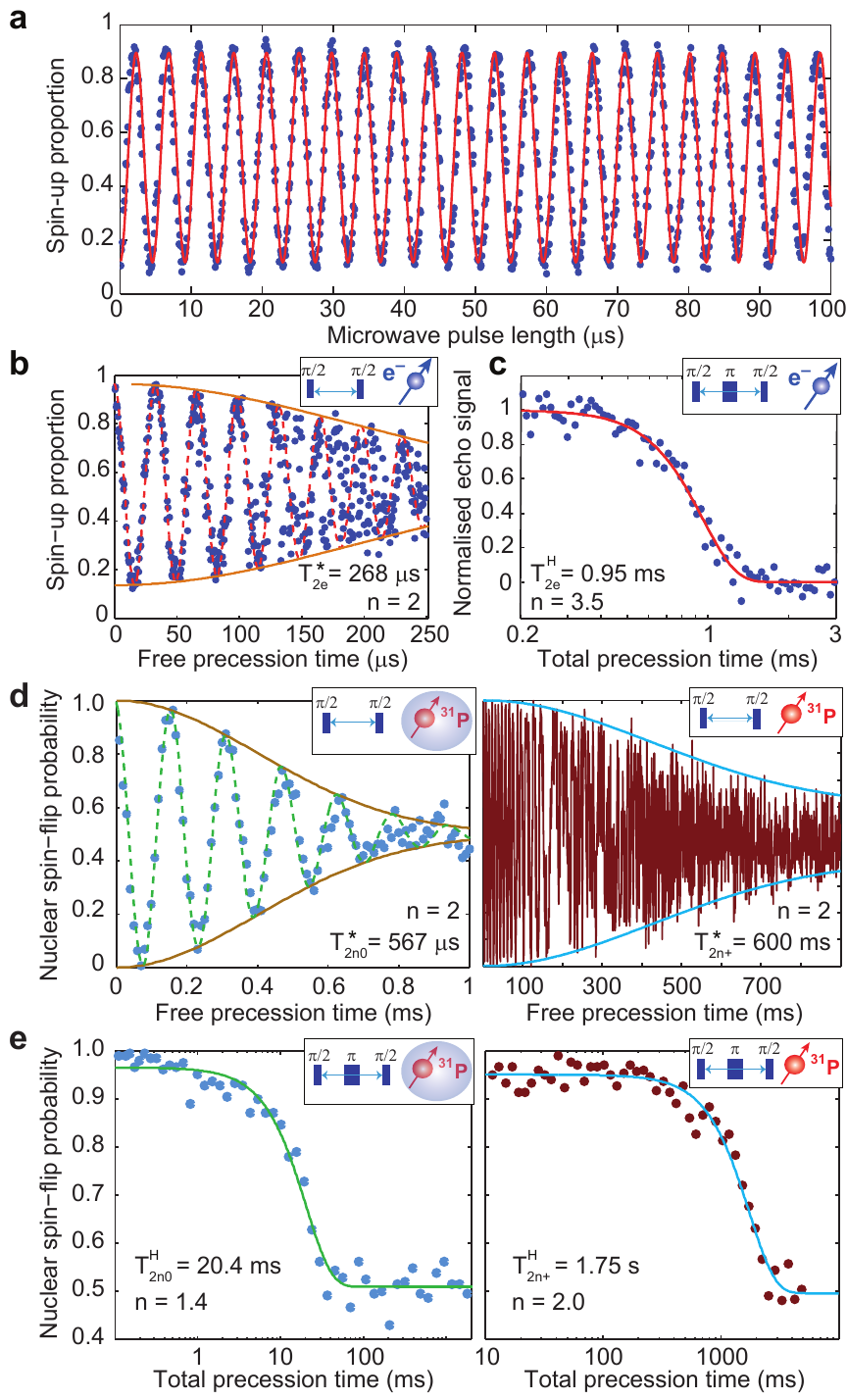}
\caption{\textbf{Coherent qubit operation}. \textbf{a}, Long-lasting Rabi oscillations of the e$^-$ qubit, showing no decay up to 100~$\mu$s. \textbf{b}, e$^-$ Ramsey fringes. \textbf{c}, e$^-$ Hahn echo decay. \textbf{d} Nuclear qubit Ramsey fringes in the neutral (left) and ionized (right) state. \textbf{e}, Nuclear Hahn-Echo decays in the neutral (left) and ionized (right) state. The coherence times quoted in each panel are obtained by fitting the envelope decays with functions of the form $P_0\exp(-(t/T_2)^n)+P_{\infty}$. The decay exponent $n$ is related to the frequency dependence of the power spectral density $S(\omega)$ of the noise that couples to the qubit (see supplementary section E). The electron Hahn-echo plot is normalized with respect to $P_0$ and $P_{\infty}$.
}
\label{fig:fig2}
\end{figure}

The qubit measurement fidelities $F_m$ were extracted from the data in Fig.~3, using a method developed in earlier work \cite{Pla2012,Pla2013}. For the e$^-$ qubit, $F_m$ is limited by the interplay of measurement bandwidth and electron tunnel times \cite{Morello2010}, and by the occurrence of false spin-up counts due to thermal effects. Through careful filtering of the signal lines we reduced the electron temperature to $\approx 100$~mK, and achieved a measurement fidelity $F_{m} \approx 97$\%. For the $^{31}$P qubit, the readout fidelity depends on the ratio between the readout time and the average time between spin flips \cite{Pla2013}. Here we achieved $F_m \approx 99.99$\%.

The use of isotopically purified $^{28}$Si brought a dramatic improvement in the qubit control fidelities. In $^{nat}$Si, the e$^-$ control fidelity was limited to $F_c = 57$\% \cite{Pla2012} by the randomness of the instantaneous resonance frequency, which fluctuates over a range comparable to the spectral width of the control pulse. Here instead the ESR linewidth is two orders of magnitude smaller than the excitation pulse spectrum, which would yield an intrinsic control fidelity of order 99.9999\%. Therefore the control errors arise solely from variation in pulse parameters due to technical limitations of the room-temperature electronic set-up. The latter can be estimated by comparing the coherence decay obtained from CPMG, which is insensitive to pulse errors up to fourth order, and from Carr-Purcell (CP), where the errors accumulate \cite{Morton2005}. With this method we obtained effective control fidelities $F_c^e \approx 99.6$\% for e$^-$, 99.9\% for $^{31}$P$^0$ and 99.99\% for $^{31}$P$^+$  (data plots presented in supplementary section D).

\begin{figure}
\centering
\includegraphics[width=0.47\textwidth]{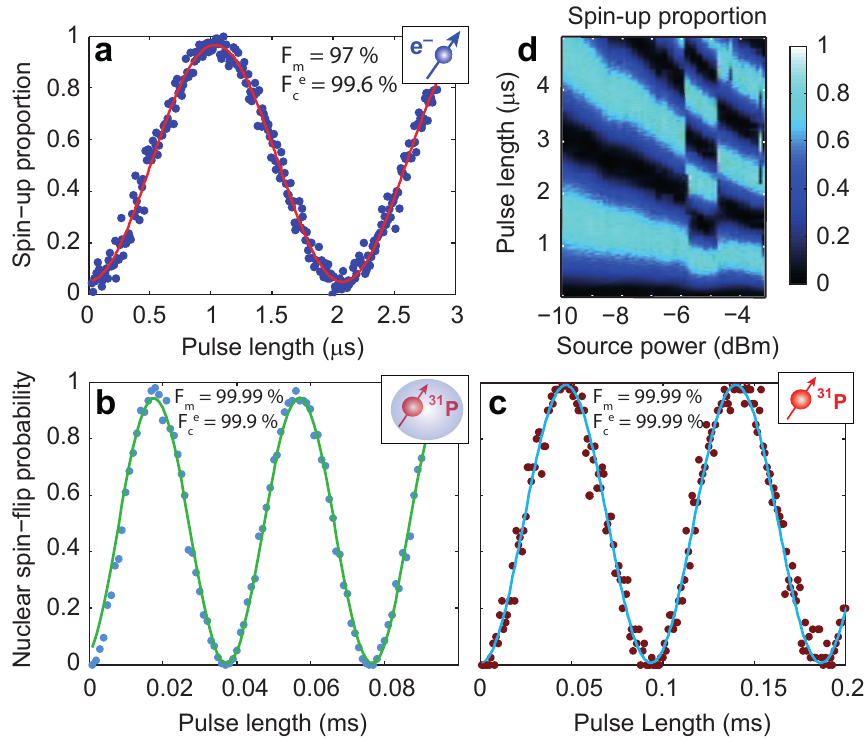}
\caption{\textbf{Rabi oscillations and qubit fidelities}. High-resolution Rabi oscillations for the electron (\textbf{a}), the neutral $^{31}$P nucleus (\textbf{b}) and the $^{31}$P nucleus (\textbf{c}). From these data we extracted the measurement ($F_m$) fidelities quoted in the figures. The effective control fidelities ($F_c^e$) are obtained by comparing CP and CMPG coherence decays (supplementary section D). \textbf{d}, 2D plot of e$^-$ Rabi oscillations at varying microwave powers. The jumps in the plot are caused by flips of the $^{31}$P nucleus between $|{\Uparrow}\rangle$ and $|{\Downarrow}\rangle$. Rabi nutation pulses are applied at both $\nu_{e1}$ and $\nu_{e2}$, but the resulting Rabi period is different in the two cases because of small variations in the frequency response of the microwave antenna and the transmission line.
}
\label{fig:fig3}
\end{figure}

Despite the record coherence times discussed above, our results do not match those obtained in bulk ensembles \cite{Tyryshkin2012,Steger2012, Saeedi2013}. We investigate the microscopic origin of spin decoherence in our nanoelectronic device by performing a systematic analysis of the spectral properties of the noise power $S(\omega)$ that modulates the e$^-$ qubit energy splitting. We adopt a noise spectroscopy method based on the properties of CPMG sequences, which act as a band-pass filter for the noise \cite{Bylander2011} with passband frequency centered at $\omega_p = \pi/\tau$, where $\tau$ is the delay between the $\pi$-pulses (see supplementary section E for details). Therefore, by choosing different $\tau$ we shift the center frequency of the filter, i.e. which portion of the noise spectrum couples the qubit. The benefits of dynamical decoupling are easily understood by considering a colored noise, e.g $S(\omega) \propto 1/\omega$. Adding more $\pi$-pulses, thus reducing $\tau$, shifts $\omega_p$ to higher frequency where the noise is weaker. For the same reason, dynamical decoupling is ineffective in the presence of frequency-independent (white) noise.

In Fig.~4\textbf{b} we show $S(\omega)$ extracted using the method described in Ref.~\cite{Alvarez2011}, which accounts for the higher harmonics in the CPMG filter function, giving small corrections to the simple relation $S(\omega) = \pi^2/(4T_2^{\rm S})$ that would hold when considering the first harmonic only (simple band-pass filter). Here $T_2^S$ is the measured electron coherence time when keeping $\tau$ constant and progressively increasing the number of pulses in a CPMG sequence. At frequencies $\omega/2\pi > 3$~kHz the noise spectrum appears flat, $S(\omega) \approx 10$~(rad/s)$^2$/Hz in Device A, corresponding to $T_2^{\rm S} \approx 0.2$~s. (For white noise, summing all the harmonics of the filter function leads to $S(\omega) = 2/T_2^{\rm S}$.)
Assuming that the noise is of magnetic origin, this corresponds to a longitudinal magnetic field noise $b_n = \hbar \sqrt{S(\omega)}/(g \mu_B) = 18$~pT/$\sqrt{\rm{Hz}}$. It's interesting to notice that substituting the simple band-pass formula here, we would recover the equation for sensitivity obtained by viewing the e$^-$ qubit as an a.c. magnetic field sensor $\eta_{a.c.} = \pi \hbar / (2 g \mu_B \sqrt{T_2})$ \cite{Taylor2008}.

Through a finite-elements modeling of the magnetic field produced by the microwave antenna at the qubit location (see supplementary section F), we calculate that this noise amplitude would correspond to the Johnson-Nyquist thermal noise produced by a 76~$\Omega$ resistor at 300 K. This is remarkably close to the 50~$\Omega$ output impedance of the MW source. Having identified the source of broadband noise that limits the ultimate qubit coherence, we added a further 7 dB attenuation at the 1.5~K stage of our dilution refrigerator before measuring Device B. This has the effect of reducing the amount of room-temperature thermal radiation that reaches the qubit. As expected, Device B exhibits a reduced white-noise floor, $S(\omega) \approx 6$~(rad/s)$^2$/Hz and $b_n = 14$~pT/$\sqrt{\rm{Hz}}$.

Both devices exhibit a colored noise spectrum below 3 kHz, approximately $S(\omega) \propto \omega^{-2.5}$ (see Fig.~4\textbf{b,c}). We attribute this low-frequency noise to instability of the external magnetic field $B_0$, based on several arguments.
First, we rule out magnetic noise from other paramagnetic spins or defects, on the basis that at $T=100$~mK and $B_0=1.5$~T any paramagnetic center is fully polarized, and its spin fluctuations exponentially suppressed \cite{Desousa2007}. This constitutes the main difference between our work and earlier bulk experiments \cite{Schenkel2006}.
Second, we rule out decoherence from locally fluctuating charges, e.g. two-level traps at the Si/SiO$_2$ interface. We have verified that the qubit is sensitive to electric field noise, by repeating the noise spectroscopy experiment in the presence of an oscillating voltage at 5~kHz, applied to an electrostatic gate above the qubit location (see supplementary section F). This sensitivity could arise from Stark-shift of the hyperfine coupling \cite{Rahman2007,Mohiyaddin2013} or the $g$-factor \cite{Rahman2009}. However the gate voltage amplitudes necessary to observe an effect in $S(\omega)$ are orders of magnitude larger than the charge noise we would expect in our device. Moreover, we have used the SET to measure the spectrum of the intrinsic charge noise in the device (see supplementary section F), and found it to follow the expected $1/\omega^\alpha$ dependence with $\alpha \approx 0.5$. This is in stark contrast with the measured spectrum of the noise acting on the qubit with $\alpha=2.5$.
Thirdly, devices A and B exhibit nearly identical low-frequency noise, whereas the SiO$_2$ in our devices has a typical trap density $\sim 10^{10}$/eV/cm$^{2}$ \cite{Johnson2010}, i.e. an average of one trap every $\sim 100$~nm. It is therefore extremely unlikely that donors in two different devices should couple to the same fluctuating charge environment. Conversely, noise from the external magnet would obviously appear with the same strength in both devices, measured with the same setup. In any case, the main message we learned from the noise spectroscopy is that the \emph{ultimate} limit to the qubit coherence, as obtained with dynamical decoupling, is set by thermal noise from the microwave antenna, and not by any noise processes intrinsic to the nanoelectronic device.

\begin{figure*}
\centering
\includegraphics{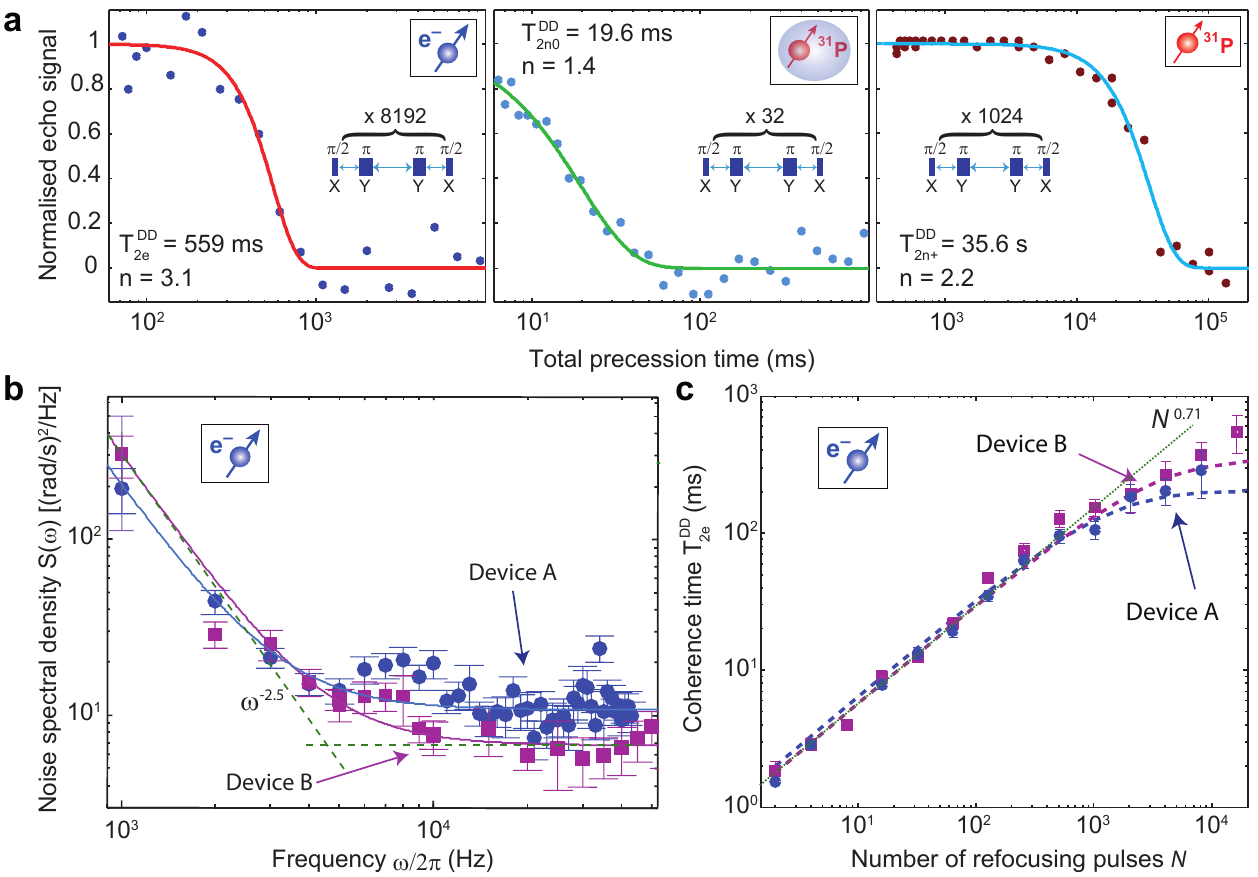}
\caption{\textbf{Coherence limits and noise spectroscopy.} \textbf{a}, Qubit coherence times obtained with CPMG dynamical decoupling sequences for electron (left), neutral nucleus (middle) and ionized nucleus (right). The number of pairs of $\pi$-pulses is indicated in the insets. \textbf{b}, Noise power spectral densities for Device A (dots), and Device B (squares). Solid lines are fits of the form $C_2/\omega^{2.5}+C_0$, with the following parameters: $C_2 = 6 \times 10^{11}$, $C_0 = 10$ for Device A; $C_2 = 9\times 10^{11}$, $C_0 = 6$ for Device B. The dashed lines show the two terms separately for device B.
 \textbf{c}, Electron coherence times $T_{2e}^{\rm DD}$ from CPMG pulse sequences as a function of the number $N$ of refocusing pulses. Lines are theoretical predictions assuming the noise spectral density from the fits in b. The excellent agreement between the calculated lines and the data at low $N$ proves that the S($\omega) \propto  \omega^{-2.5}$ dependence continues well below the 1 kHz measured in b.
 With $S(\omega) \propto \omega^{-\alpha}$, the coherence time should be proportional to $N^{\alpha/(\alpha+1)}$ , yielding $N^{0.71}$ for $\alpha = 2.5$, which is shown in the figure as a guide to the eye.
Error bars are 95\% confidence intervals from the exponential fits used to extract the decay times.
}
\label{fig:fig4}
\end{figure*}

Our results conclusively show that the exceptional quantum coherence exhibited by spins in isotopically pure $^{28}$Si can be preserved and exploited in a top-gated nanoelectronic device, fabricated with standard Metal-Oxide-Semiconductor methods. The proximity of an amorphous interface and gated nanostructures does not appear to significantly affect the control fidelity and the coherence time, which reaches here a new record for solid-state single qubits with $T_2 > 30$~s in the $^{31}$P$^+$ spin. Looking beyond the single-qubit level, we note that the most promising proposals for 2-qubit logic gates and long-distance coupling involve rather weak interactions, either through exchange coupling \cite{Kalra2013}, or in a circuit quantum electrodynamics (cQED) architecture \cite{Hu2012}. The extremely narrow linewidths observed here will facilitate multi-qubit operations based on magnetic resonance, since the individual resonances will remain resolvable over a very broad range of inter-qubit couplings, greatly relaxing the need for atomically precise donor placement. This work represents a fundamental advance in control and understanding of spin qubits in solid state, and and shows a clear path forward to integrating them with functional electronic devices.

\section*{Methods}

\subsection*{Device fabrication}
The device was fabricated on a 0.9 $\mu$m thick epilayer of isotopically purified $^{28}$Si, grown on top of a 500 $\mu$m thick $^{nat}$Si wafer. The $^{28}$Si epilayer contains 800 ppm residual $^{29}$Si isotopes. Single-atom qubits were selected out of a small group of donors implanted in a region adjacent the Single-Electron-Transistor. In Device A, individual P$^+$ atoms were ion-implanted at 14 keV energy in a $90 \times 90$~nm$^2$ window defined by a resist mask. In Device B, P$_2^+$ molecular ions were implanted at 20 keV energy, again in a $90 \times 90$~nm$^2$ window. All other nanofabrication processes were identical to those described in detail in Ref. \cite{Pla2012}, except for a slight modification in the gate layout to bring the qubits closer to the microwave antenna and provide an expected factor 3 improvement in $B_1$.

\subsection*{Experimental Setup}
The sample was mounted on a high-frequency printed circuit board in a copper enclosure, thermally anchored to the cold finger of an Oxford Kelvinox 100 dilution refrigerator with a base temperature $T_{bath}=20$~mK. The sample was placed in the center of a wide-bore superconducting magnet, oriented so the $B_0$ field was applied along the [110] plane of the Si substrate, and perpendicular to the short-circuit termination of the MW antenna. The magnet was operated in persistent mode while also feeding the nominal current through the external leads. We found that removing the supply current while in persistent mode led to a very significant magnetic field and ESR frequency drift, unacceptable given the intrinsic sharpness of the resonance lines of our qubit. Conversely, opening the persistent mode switch led to noticeable deterioration of the spin coherence, most visible as a shortening of $T_2^*$ in Ramsey experiments.

Room-temperature voltage noise was filtered using an anti-inductively wound coil of thin copper wire with a core of Eccosorb CRS-117 ($\sim 1$~GHz cut-off), followed by two types of passive low-pass filters: 200 Hz second-order $RC$ filters for DC biased lines, and 80 MHz seventh-order Mini-Circuits $LC$ filters for pulsed voltage lines. The filter assemblies were placed in copper enclosures, filled with copper powder, and thermally anchored to the mixing chamber.
DC voltages were applied using optoisolated and battery-powered voltage sources, connected to the cold filter box via twisted-pair wires. Voltage pulses were applied using an arbitrary waveform generator (LeCroy ArbStudio 1104), connected to the filter box via semi-semirigid coaxial lines.

ESR pulses were generated using an Agilent E8257D analog signal generator, and NMR pulses were produced by an Agilent MXG N5182A vector signal generator. Both excitation signals were combined using a power-combiner and fed to the MW antenna via a CuNi semi-rigid coaxial cable, with attenuators at the 1.5 K stage (3 dB for Device A, 10 dB for Device B) and the 20 mK stage (3 dB). The pulse and phase modulation of both microwave sources was controlled using a PCI TTL pulse generator (SpinCore PulseBlaster-ESR). The SET current was measured by a Femto DLPCA-200 transimpedance amplifier at room temperature, followed by a floating-input voltage post-amplifier, a sixth-order low-pass Bessel filter, and acquired using a PCI digitiser card (AlazarTech ATS9440).

\subsection*{Data acquisition statistics}
For e$^-$ experiments the state is always initialized spin-down and all of our plots were produced by taking the spin-up proportion from 100-200 single-shot measurement repetitions per point. For $^{31}$P experiments, plots were produced by taking the nuclear flipping probability (no initialization to a certain state) from 41 measurement repetitions per point, and 50 electron readouts per nuclear readout. See Ref. \cite{Pla2013} for more details on nuclear readout and control sequences.

\textbf{Acknowledgments} This research was funded by the Australian Research Council Centre of Excellence for Quantum Computation and Communication Technology (project number CE11E0096) and the US Army Research Office (W911NF-13-1-0024). We acknowledge support from the Australian National Fabrication Facility, and from the laboratory of Prof Robert Elliman at the Australian National University for the ion implantation facilities. The work at Keio has been supported in part by FIRST, the Core-to-Core Program by JSPS, and the Grant-in-Aid for Scientific Research and Project for Developing Innovation Systems by MEXT.
\newline \\
\textbf{Authors Contributions} J.T.M., J.P.D., A.S.D. and A.M. designed the experiments. J.T.M., J.P.D. and A.L. performed the measurements and analysed the results with A.M.'s supervision. D.N.J. and J.C.M. designed the P implantation experiments. F.E.H. fabricated the device with A.S.D.'s supervision. T.S. and K.M.I. prepared and supplied the $^{28}$Si epilayer wafer. J.T.M., J.P.D and A.M. wrote the manuscript, with input from all coauthors.
\newline \\
The authors declare no competing financial interests.
\newline \\
Correspondence should be addressed to J.T.M. (juha.muhonen@unsw.edu.au) or A.M. (a.morello@unsw.edu.au).
\newline \\
\textbf{Supplementary Information} accompanies the paper. 

\clearpage
\onecolumngrid

\renewcommand{\thefigure}{S\arabic{figure}}
\renewcommand{\thetable}{S\arabic{table}}

\section*{Supplementary Information for \\ "Storing quantum information for 30 seconds in a nanoelectronic device"}

\subsection{Summary of the measured coherence times}

\begin{table}[h!]
\centering
\includegraphics{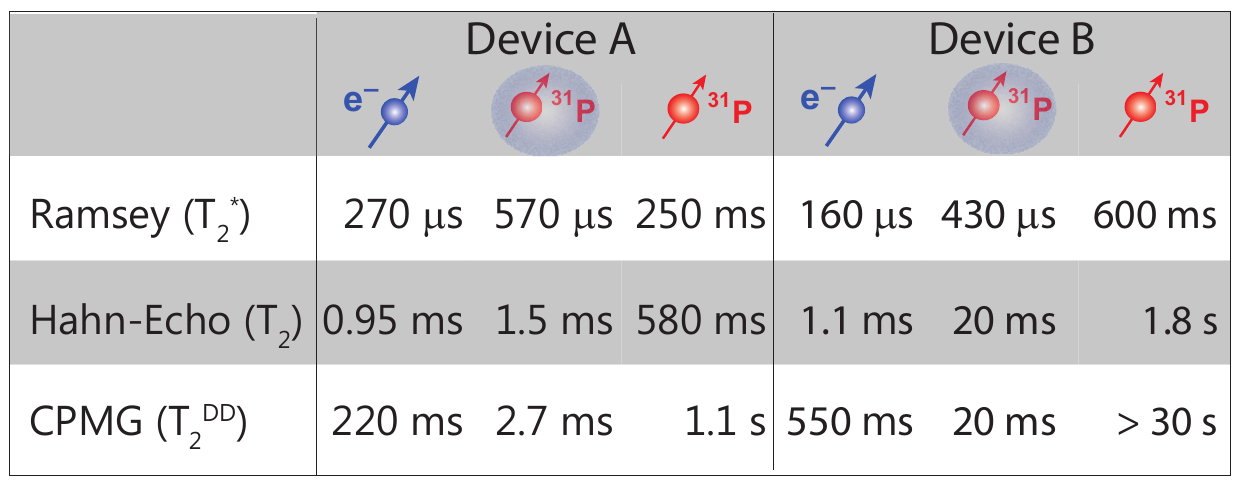}
\caption{\textbf{Coherence limits.} Summary table of the coherence benchmarks for the electron (e$^-$), the neutral nucleus ($^{31}$P$^0$) and ionized nucleus ($^{31}$P$^+$) qubits. Only significant difference between devices A and B is that the thermalization of the microwave line was improved for device B. The ionized nucleus $T_2^{DD}$ times are not fully comparable as with Device A the number of DD pulses was not pushed to the limits in this case.}
\end{table}

\subsection{Pulse sequences for qubit operation}

The performance of the qubits was measured using the standard magnetic resonance techniques. We define a rotating reference frame where $Z$ is the direction of the static $B_0$ field, while $X$ and $Y$ rotate around $Z$ at the qubit precession frequency, so they appear static in the reference frame of the qubit. A rotation of the spin by an angle $\theta$ around the $\Phi$ axis in the rotating reference frame is indicated as $\Phi_{\theta}$. By convention, $X$ is defined by the phase of the first resonant pulse applied to the qubit, whereas a pulse around $Y$ is obtained through a $90^{\circ}$ phase shift of the microwave source. Arbitrary quantum states are encoded on the qubits by applying pulses of oscillating magnetic field $B_1$ at the frequencies corresponding to the electron spin resonance (ESR), $\nu_{e1,2} \approx \gamma_e B_0 \pm A/2$, and nuclear magnetic resonance (NMR), $\nu_{n1,2} \approx A/2 \pm \gamma_n B_0$, where $\gamma_e = 27.97$~GHz/T and $\gamma_n = 17.23$~MHz/T are the electron and nuclear gyromagnetic ratios, respectively, and $A \approx 117$~MHz is the hyperfine coupling. The $^{31}$P qubit can also be operated at the frequency $\nu_0 = \gamma_n B_0$ while the electron is absent, i.e. the P atom is ionized \cite{SPla2012,SSaeedi2013}.

All of the measurements described in the main text consist of two phases: a control phase and a readout phase. During the control phase, the donor potential is tuned below (for e$^-$ and $^{31}$P$^0$) or above (for $^{31}$P$^+$) the Fermi level of the SET island, to maintain the donor in the neutral or ionized state respectively, during the application of control pulses. The readout phase follows directly after the control phase and consists of a single-shot electron readout \cite{SMorello2010} (for e$^-$), or a single-shot nuclear readout \cite{SPla2013} (for $^{31}$P). Each measurement is repeated several times to compute the spin-up proportion $P_\uparrow$ (for the electron) or the spin flip probability $P_f$ (for the nucleus).

We applied the following pulse sequences:

\textbf{(i) Rabi oscillations}, obtained by monitoring $P_{\uparrow}$ (for e$^-$) or $P_f$ (for $^{31}$P) as a function of the duration $\tau_{R}$ of a pulse $X_{\theta}$, with $\theta = 2\pi \tau_R / \gamma B_1$;

\textbf{(ii) Ramsey fringes}, obtained by applying a $X_{\pi/2}$ pulse, followed by a free precession time $\tau$, then another $X_{\pi/2}$ pulse-pulse that brings the spin back along the $Z$-axis for measurement ($X_{\pi/2} - \tau - X_{\pi/2}$). $P_{\uparrow}$ or $P_f$ oscillate if the frequency of the microwave source is detuned from the qubit frequency, and the decay of the oscillations' envelope yields the pure dephasing time $T_2^*$;

\textbf{(iii) Hahn echo}, obtained by introducing a $X_{\pi}$ pulse between the $X_{\pi/2}$ pulses in a Ramsey sequence ($X_{\pi/2} - \tau/2 - X_{\pi} - \tau/2 - X_{\pi/2}$). The $X_{\pi}$ pulse cancels the effect of random variations of the instantaneous qubit frequency that are static over the timescale of a single experimental run, and yields the qubit coherence time $T_2^{\textrm{H}}$;

\textbf{(iv) Carr-Purcell-Meiboom-Gill (CPMG)}, where the $X_{\pi}$ pulse in the Hahn echo sequence is replaced by $N$ $Y_{\pi}$ pulses separated by $\tau$ ($X_{\pi/2}-(\tau/2-Y_{\pi}-\tau/2)^N-X_{\pi/2}$). This sequence is often used to extend the timescale over which a quantum coherent state can be preserved. The $Y_{\pi}$ pulses make it first-order immune to imperfections in the pulse lengths. The decay measurement can be performed in two ways: by fixing $N$ and varying $\tau$, or by fixing $\tau$ and varying $N$. The measurements differ in how the effective noise filter changes after each sequence increment. Here we use the latter sequence to extract the spectrum of the noise that couples to the qubit, and the first sequence to measure the ultimate limit of $T_2$. A variant of CPMG is the Carr-Pucell (CP) sequence, where the $Y_{\pi}$ pulses are replaced with $X_{\pi}$. This sequence has the same filter function as CPMG, but loses the immunity to pulse errors. Here we use this sequence to extract the control fidelity of our qubits.

In each case, the qubit coherence decays as a function of the total wait time $t$ with a law of the form $P(t) = P_0 \exp\left[-\left(t / T_2 \right)^n\right] + P_{\infty}$. The decay exponent $n$ is related to the frequency dependence of the power spectral density $S(\omega)$ of the noise that couples to the qubit (see noise spectroscopy section).

\subsection{Electron spin resonance linewidths}

The free induction decay time ($T_2^*$) is intrinsically related to the full-width half-maximum of the ESR line, $\Delta\nu_{\textsc{fwhm}} = 1 / (\pi T_{2e}^*)$. To measure the linewidth directly in an ESR spectrum experiment, the excitation profile of the applied microwave must be narrower than $\Delta\nu_{\textsc{fwhm}}$, otherwise the excitation spectrum will dominate the measurement. In continuous-wave experiments the excitation profile is generally very narrow, however our readout method requires the excitation to be pulsed, which makes it difficult to achieve very narrow excitation spectra.

\begin{figure}
\centering
\includegraphics{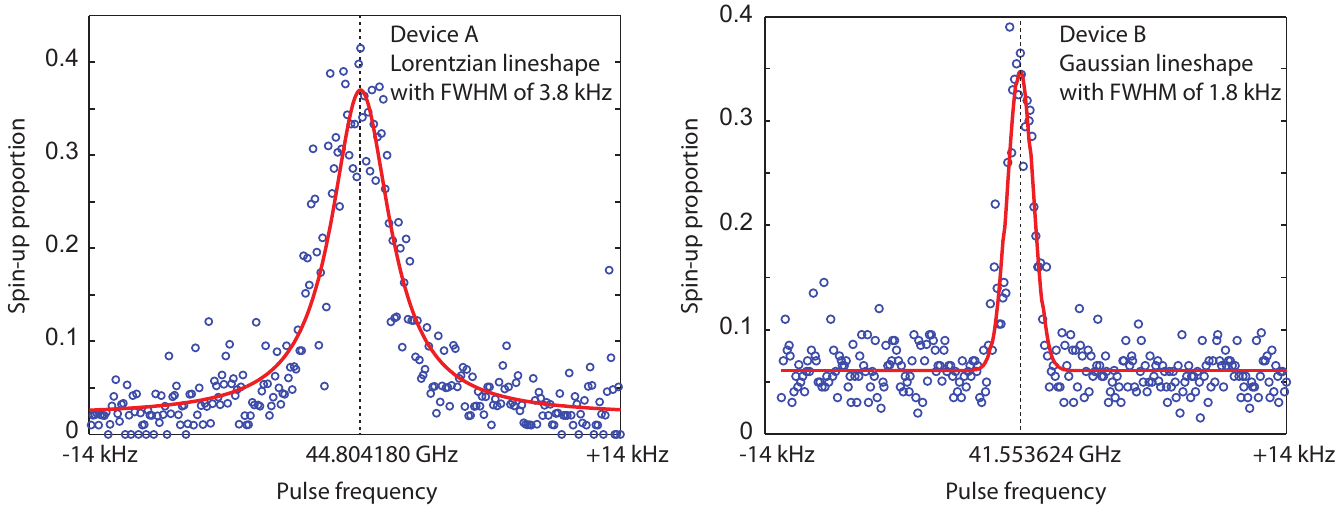}
\caption{\textbf{Direct measurement of electron spin resonance linewidths.} The ESR measurements were performed with -50 dBm of MW source power and 500 $\mu$s pulse width. Device A was measured using square--shaped pulses, while device B was measured using Gaussian--shaped pulses. Solid lines are a Lorentzian fit (Device A) and a Gaussian fit (Device B), with parameters as quoted in the figures.}
\label{fig:figS1}
\end{figure}

One method commonly used to narrow the excitation profile in pulsed experiments is to apply pulse shaping. For device A, we used square pulses at -50 dBm of source power, obtaining a linewidth of 3.8~kHz with a Lorentzian shape, which suggests that it is still power broadened. On device B we performed the same experiment but used Gaussian-shaped pulses, which allowed us measure a $\Delta\nu_{\textsc{fwhm}} = 1.8$~kHz, much closer to the intrinsic linewidth. The lineshape here was Gaussian, consistent with the $n=2$ exponent of the Ramsey decay. The measured ESR lines are shown in Fig.~\ref{fig:figS1}.

The values of $\Delta\nu_{\textsc{fwhm}}$ we observed are substantially smaller than those measured in bulk ensembles, even in ultra-pure $^{28}$Si sourced from the Avogadro Project \cite{STyryshkin2012} which contain $< 50$~ppm residual $^{29}$Si. With 800 ppm residual $^{29}$Si in our epilayer, the expected number of $^{29}$Si nuclei in the 2.5~nm Bohr radius of the electron wave function is less than 3. This brings us in a very peculiar regime where the `spin bath' is a small and discrete system, and comparisons with ensemble-averaged experiments are not meaningful. In addition, a single-atom experiment is intrinsically immune from inhomogeneities in the $g$-factor and the hyperfine coupling. The $n = 2$ decay exponent of the Ramsey oscillations corresponds to a Gaussian lineshape in the frequency domain, as expected for a single spin in a dilute spin bath \cite{SDobrovitski2008}.

\subsection{Control and measurement fidelities}

\begin{figure}
\centering
\includegraphics{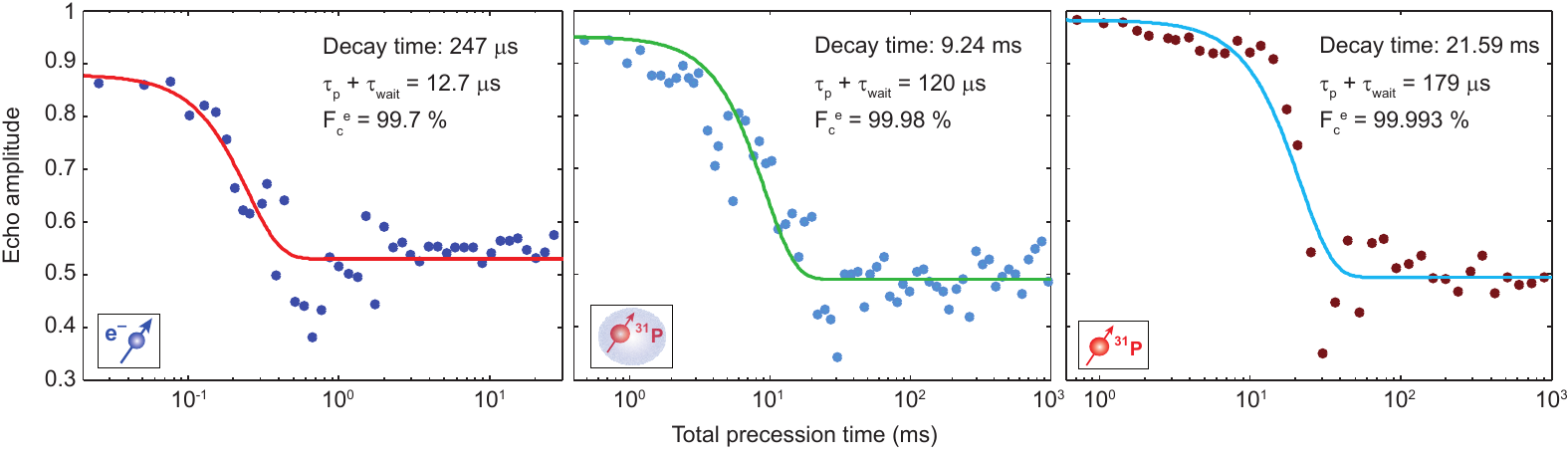}
\caption{\textbf{Carr-Purcell (CP) measurements to extract the qubit control fidelities.} The CP echo decays were measured on the electron (left), the neutral nucleus (middle) and the ionized nucleus (right) by applying $\pi$-pulses along the $x$-axis at short intervals. In this pulse sequence, the decay is dominated by the accumulation of pulse errors. The qubit control fidelity is extracted from the fits described in the text.}
\label{fig:figS2}
\end{figure}

The control fidelities of the 3 different qubits (electron, neutral nucleus and ionized nucleus) were all extracted using the method presented in Ref.~\cite{SMorton2005}. By comparing the decays of a CPMG sequence with a CP, the pulse-error component of the decay can be extracted. Assuming Gaussian-distributed pulse errors, the decay is of the form:
\begin{equation}
P(N) \propto \exp\left(-(\sigma N/2)^2\right),
\end{equation}
where $\sigma$ is the standard variation of the rotation angle (in radians) with a mean of $\pi$, and $N$ the number of pulses. To extract the control fidelity, we first apply a CPMG sequence with a fixed $\tau_{wait}$ and extract $T_2^{\textsc{cpmg}}$ and the exponent $n$. We then apply a CP sequence with the same $\tau_{wait}$ and fit the decay to:
\begin{equation}
P(t_p) \propto \exp\left(-\left(\frac{t_p}{T_2^{\textsc{cpmg}}}\right)^n\right)\exp\left(-\left(\frac{\sigma t_p}{2(\tau_{wait}+\tau_\pi)}\right)^2\right), \label{Ptp}
\end{equation}
where $t_p = N(\tau_{wait}+\tau_\pi)$ is the total precession time between the initial and final $X_{\pi/2}$ pulses, and $\tau_\pi$ is the duration of each $\pi$-pulse. In the practice, we usually found that the value of $T_2^{\textsc{cpmg}}$ was so long that the term $\exp\left(-(t_p/T_2^{\textsc{cpmg}})^n\right)$ could be approximated with 1.
After extracting $\sigma$ from the fit of Eq.~\ref{Ptp} to the CP decay data, we define the effective control fidelity as:
\begin{equation}
F_c^e = \frac{1}{2}[\cos\left(\sigma\right)+1].
\end{equation}
Data from the CP sequences used to extract the control fidelity is shown in Fig.~\ref{fig:figS2}.

The e$^-$ measurement fidelity is defined as in ref. \cite{SPla2012}. It is extracted from the Rabi oscillation data shown in the main text Fig.~3a, and from histograms of the maximum current during readout. From the latter we extract an electrical visibility of 98\% at the optimal threshold. The overall fidelity of 97\% is limited by thermal broadening of the electron distribution on the readout SET. Our electron temperature is estimated to be $T_{el} \approx 100$~mK.

For the nuclear qubit measurement fidelity we use the same methods as in \cite{SPla2013}. As a single readout of the nuclear qubit is done by 50 single-shot readouts of the electron state, the measurement fidelity is very close to unity. The limiting factor becomes the possibility of a random nuclear flip during the relatively long readout time. The nuclear lifetimes were 2900 s for the $|{\Downarrow \rangle}$ state and 7900 s for the $|{\Uparrow \rangle}$ state.  Using the total readout time of 250 ms ($50 \times 5$ ms) we get the measurement fidelity quoted in the main text (we always quote the worst-case scenario of the shortest-lived nuclear state). All these values were measured on Device A.

\subsection{Details of noise spectroscopy}

Dynamical decoupling is a well known method in spin resonance community to cancel out low frequency noise. By applying $\pi$-pulses with regular intervals ($\tau$) one effectively averages out noise at frequencies much lower than $1/2\tau$.  For random noise with a mean of zero, noise at much higher frequencies also averages to zero.  As a result it can be very useful to think about dynamical decoupling (DD) pulsing schemes as band-pass spectral filters for the noise \cite{SUhrig2007,SCywinski2008,SBiercuk2011}. If we consider pure dephasing noise with a Hamiltonian
\begin{equation}
\mathcal{H} = \frac{\hbar}{2}\left[\Omega + \beta(t)\right]\sigma_z,
\end{equation}
where $\Omega$ is the Larmor frequency ($\hbar\Omega$ is the energy splitting of the qubit states) and $\beta(t)$ is time-dependent noise (in angular frequency units), it can be shown \cite{SBylander2011,SYuge2011,SAlvarez2011} that the decay of the coherence is of the form
\begin{equation}
P(N\tau) \propto \exp\left(-\int_{0}^\infty S(\omega) |F(\omega,N\tau)|^2 d\omega\right),
\end{equation}
where $S(\omega) = \int_{-\infty}^\infty \exp(-i\omega t) \langle\beta(t)\beta(0)\rangle dt$ is the power spectral density (PSD) of the noise and $|F(\omega,N\tau)|^2$ is a pulse sequence dependent function known as the filter function since it determines which parts of the noise spectra contribute to the decoherence process. The total evolution time is $N\tau$ where $N$ is the number of pulses. The pulses are assumed to be instantaneous. We define $S(\omega)$ as the noise power in the energy splitting of the qubit states (in angular frequency units), as opposed to making assumption on its physical nature (magnetic, electric, etc.) and adding a coupling constant in Eq. (5).

\begin{figure}
\centering
\includegraphics{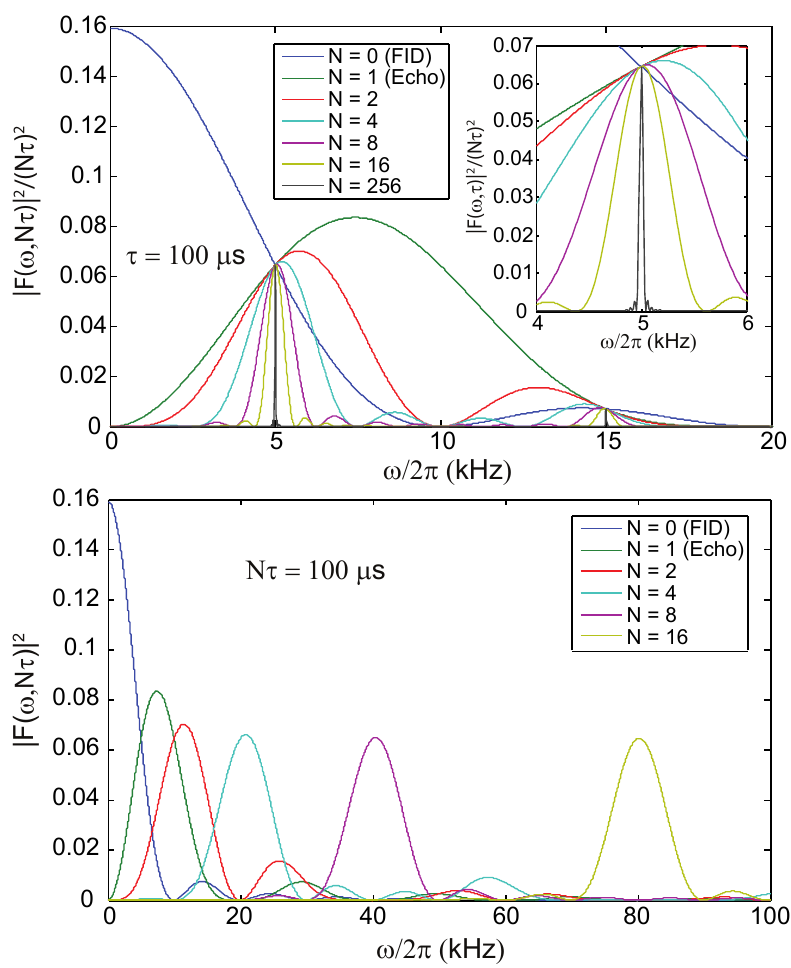}
\caption{\textbf{The filter function.} Noise filter function of the CPMG sequence, plotted with fixed interval between the pulses (top) and with fixed total evolution time (bottom). The top plot is normalized with $(N\tau)^2$, as the height of the peak grows with the square of the total evolution time. (The width of the peak scales roughly as $(N\tau)^{-1}$.) In the bottom plot no normalization is needed as the area under the curve stays constant.}
\label{fig:figS3}
\end{figure}

The filter function $F(\omega,N\tau)$ has an analytical expression for the case of $\pi$-pulses applied at regular intervals \cite{SCywinski2008}
\begin{equation}
 |F(\omega,N\tau)|^2 = \frac{8}{\pi} \frac{1}{\omega^2} \frac{\sin^4(\omega\tau/4)\sin^2(\omega N \tau/2)}{\cos^2(\omega\tau/2)}.
\end{equation}
This has been plotted in Fig.~\ref{fig:figS3} with a constant interval between the pulses (top) and with a constant total evolution time (bottom). The top figure is relevant for the noise spectroscopy (Fig.~4b in the main text) whereas the bottom one describes how the passband frequency moves with increasing $N$ at constant total evolution time (Fig.~4c in the main text).

Notably, if one keeps the total free evolution time of the qubit constant, the integral over the filter function ("the bandwidth") is the same for all regular interval $\pi$-pulsing schemes. Hence, if the dephasing noise has a fully flat (white) spectral density, the refocusing pulses will make no difference to the decoherence time, whereas for strongly frequency dependent noise DD can be very effective. The powerlaw of the exponential decay will also depend on the shape of the dephasing noise within the "band" of the filter, giving exponential decay ($n=1$) for white noise, gaussian decay ($n=2$) for $ 1/\omega$ noise and so on, i.e., $S(\omega) \propto \omega^{-(n-1)}$.

It was recently pointed out \cite{SYuge2011,SAlvarez2011} that in the limit where N is large, the filter function becomes a series of delta functions and the coherence decay has a simple analytic form
\begin{equation}
P(t) \propto \exp \left(-t \frac{4}{\pi^2}\sum_{k=0}^\infty \frac{1}{(2k+1)^2}S(\omega_k)\right),
\end{equation}
where $\omega_k = (2k+1) \pi/\tau$. Hence, as proposed in \cite{SAlvarez2011} measuring the decay time $T_2^S$  at multiples of some minimum frequency $\omega_0$ (1 kHz in our case) allows one to map out the noise spectral density including the higher harmonics of the filter function. We note that taking only the first term of Eq. (7) leads to the simple band-pass filter form whereas in the case of frequency independent noise Eq. (7) reduces to $P(t) \propto \exp \left(-tS/2\right)$.

\subsection{Quantitative analysis of the noise}

We have performed extensive modeling on the electric and magnetic fields induced by the broadband antenna. The MW antenna is designed to produce no electric fields -- hence the short-circuit termination of the co-planar stripline conductors. In the practice, imperfections in the propagation along the antenna may result in nonzero electric fields at its termination. The electric field couples to the qubit energy levels by Stark-shifting the hyperfine constant $A$ \cite{SBradbury2006,SRahman2007,SMohiyaddin2013} and/or the $g$-factor of the electron spin \cite{SRahman2009}. The antenna should also produce purely transverse oscillating magnetic field $B_1 \perp Z$ , since longitudinal (i.e. $\parallel B_0$) component of the magnetic field will directly modulate the qubit Larmor frequency and cause decoherence. However, our device geometry does allow for some nonzero longitudinal field component.

\begin{figure}
\centering
\includegraphics{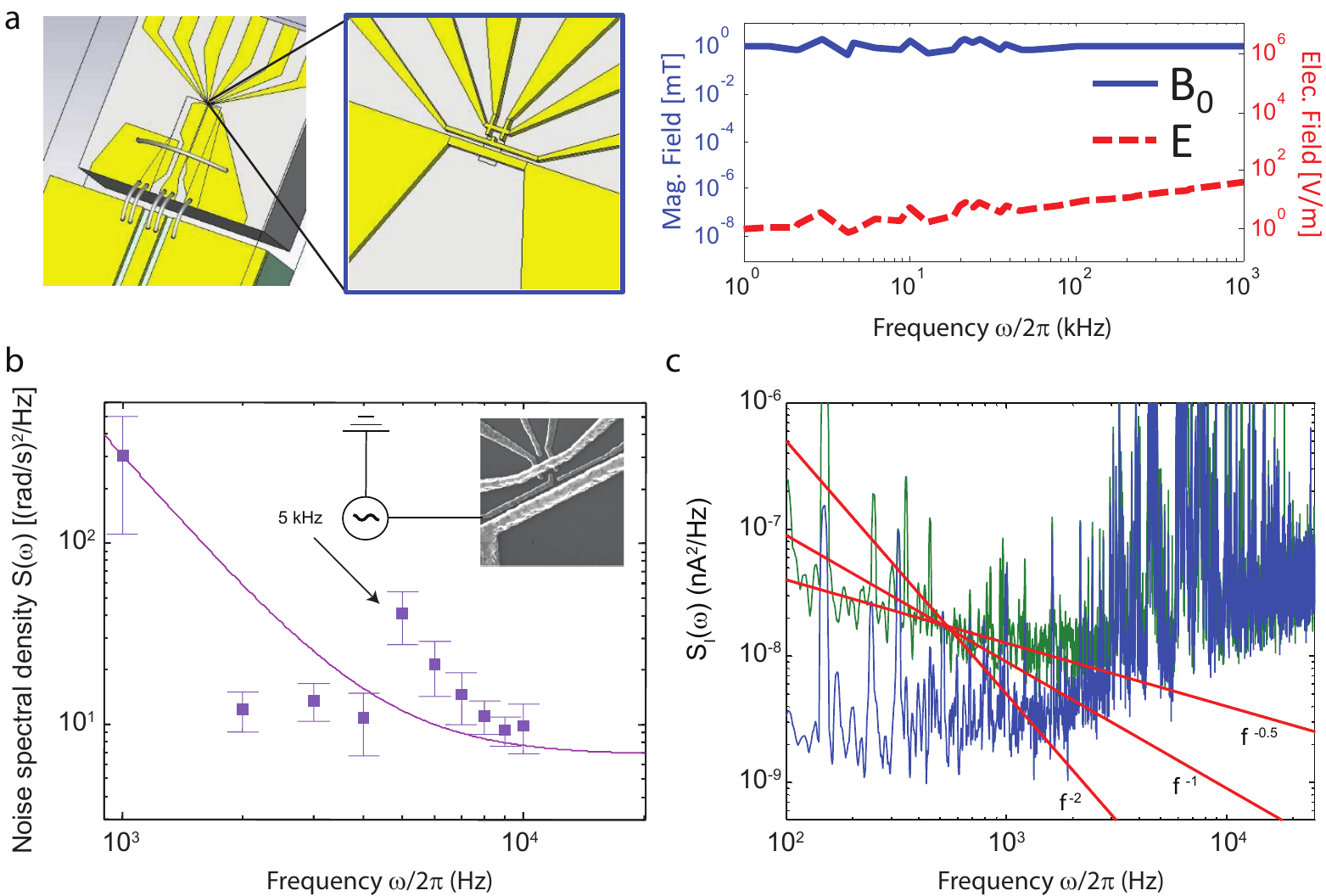}
\caption{\textbf{Analysis of noise sources.} \textbf{a,} The left drawing shows the geometry used for the MW antenna. The simulation results are shown in the right for the longitudinal magnetic field component ($B_0$) and electric field ($E$), at low frequencies. All plots are calculated by assuming an input signal of 0 dBm at each modeled frequency. Calculations of the magnetic and electric fields produced by the MW antenna were performed using a the finite-elements electromagnetic solver CST Microwave Studio.
\textbf{b,} Noise spectroscopy measurement while applying a 5 kHz voltage signal to the gate shown in the SEM image. The noise signal was a sine-wave with 100 $\mu$V amplitude (200 $\mu$V peak-peak). A clear peak appeared in the noise spectra at 5 kHz. Solid line is the same as plotted in Fig.~4\textbf{b} of the main text.
\textbf{c,} Power spectral density of current traces taken while the SET was tuned to be either maximally sensitive to charge fluctuations (green, top) or minimally sensitive (blue, bottom). The data has been numerically smoothed. The difference at low frequencies is from the charge noise in our device and follows $1/\omega^\alpha$ dependency, with $\alpha\approx0.5$. Lines are guides to the eye. At higher frequencies (above $\sim 2$ kHz) the noise is dominated by the room-temperature transimpedance amplifier and the two curves coincide. 
}
\label{fig:figS4}
\end{figure}

For the modeling we assume a donor location in the middle of the implant window, 10 nm below the Si/SiO$_2$ interface, and model the electric and magnetic fields at that location using CST Microwave Studio finite-elements software. The simulation results are shown in Fig.~\ref{fig:figS4}\textbf{a}.

For device A we measured a noise floor of $S(\omega) \approx 10$~rad$^2$/s, which converted to amplitude spectral density and to magnetic field gives $b_n = \hbar \sqrt{S(\omega)} / (g \mu_B) = 18$~pT/$\sqrt{\textrm{Hz}}$ for $g=2$. The finite-elements modeling shows that, at $\sim \textrm{kHz}$ frequencies, 1 mW of power result in a longitudinal magnetic field component $B_z \approx 1$~mT. Since the magnetic field is proportional to the square root of the power, we deduce that the noise power $P_n$ entering the MW antenna to produce $b_n = 18$~pT/$\sqrt{\textrm{Hz}}$ is:
\begin{equation}
P_n = \left(\frac{b_n}{1~\textrm{mT}}\right)^2 \times 1~\textrm{mW} = 3.2 \times 10^{-19} \textrm{W} = - 155~\textrm{dBm}
\end{equation}
During the measurement of Device A we had a total of 6 dB attenuation along the signal line (the losses of the coaxial cable are negligible at kHz frequencies), thus -155~dBm at the chip corresponds to $P_n^s = -149$~dBm noise power at the source. The power radiated down by the attenuators themselves would give maximum 2\% correction to this value as they are thermalised to the pot (1.5 K) and mixing chamber (0.02 K). If we assume that the noise power is Johnson-Nyquist noise produced by a resistor $R$ at $T = 300$~K, $P_n^s = 4 k_B T R$ per unit of frequency, we find:
\begin{equation}
R = \frac{10^{-14.9} \times 10^{-3}}{4 k_B T} = 76~\Omega.
\end{equation}
This value is remarkably close to the 50~$\Omega$ impedance presented by the output of the microwave source, especially considering that the exact conversion between $P_n$ and $b_n$ involves the large uncertainty of the donor location.

When measuring Device B we increased the attenuation of the microwave line by 7 dB, and observed accordingly a decrease of the white noise floor, albeit not by the numerical amount expected on the basis of the additional attenuation. This could be due to a different donor location in Device B, such that the $B_z$ component is larger than for device A.

To confirm the noise spectroscopy method, and to verify that the e$^-$ qubit is potentially sensitive to electric field noise, we repeated the measurement of $S(\omega)$ for frequencies between 1 and 10 kHz while applying a 5 kHz sine-wave voltage signal on one of the control gates fabricated above the donor implant window.  The data is presented in Fig.~\ref{fig:figS4}\textbf{b}. A clear peak in $S(\omega)$ appeared at the expected 5~kHz frequency, confirming the effectiveness of the method. However, the amplitude of the signal we had to apply in order to distinguish it from the background was of the order of 100 $\mu$V, much above what could be conceivably produced by charge fluctuations of interface traps.

Finally, in an effort to study the frequency dependence of the intrinsic charge noise in our device we measured few long traces of the output of our current amp while a) the device was tuned to a slope of a Coulomb peak where the current through the SET is maximally sensitive to any charge variations in its surroundings and b) when the SET was in a non-conductive region where it should be insensitive to charge noise. Comparing the spectra of these two traces we can extract the low frequency charge noise of our device (at higher frequencies the noise floor of the amplifier is the limiting noise). As shown in  Fig.~\ref{fig:figS4}\textbf{c} the low-frequency charge noise follows an $1/\omega^{\alpha}$ curve with $\alpha\approx 0.5$. This adds further evidence to the conclusion that our $1/\omega^{2.5}$ noise component is not due to charge noise.

\end{document}